\documentclass[prl,aps,epsfig,superscriptaddress,showpacs,twocolumn,notitlepage]{revtex4-1}
\usepackage{amsmath,amssymb,amsfonts,amstext}
\usepackage{graphicx,breakurl,color,setspace}
\usepackage[T1]{fontenc}
\usepackage[latin9]{inputenc}
\usepackage[normalem]{ulem}
\usepackage[unicode=true,pdfusetitle,pdfborder={0 0 1}]{hyperref}
\usepackage{upgreek}
\usepackage{bm}
\usepackage{mathrsfs}

\hypersetup{colorlinks,
linkcolor=blue,          citecolor=blue,        filecolor=blue,      urlcolor=blue           }

\newcommand{\bra}[1]{\langle #1 |} 
\newcommand{\ket}[1]{| #1 \rangle} 
\newcommand{\bracket}[3]{\langle #1 | #2 | #3 \rangle} 

\usepackage{times}

\graphicspath{{../figures/}}

\begin{document}

\title{Machine Learning Topological Phases with a Solid-state Quantum Simulator}

\author{Wenqian Lian}\thanks{These authors contributed equally to this work.}
\affiliation{Center for Quantum Information, IIIS, Tsinghua University, Beijing 100084, PR China}
\author{Sheng-Tao Wang}\thanks{These authors contributed equally to this work.}
\affiliation{Department of Physics, Harvard University, Cambridge, MA 02138, USA}
\affiliation{Center for Quantum Information, IIIS, Tsinghua University, Beijing 100084, PR China}
\author{Sirui Lu}\author{Yuanyuan Huang}\author{Fei Wang}\author{Xinxing Yuan}\author{Wengang Zhang}\author{Xiaolong Ouyang}\author{Xin Wang}\author{Xianzhi Huang}\author{Li He}\author{Xiuying Chang}
\author{Dong-Ling Deng}\email{dldeng@mail.tsinghua.edu.cn}
\author{Luming Duan}\email{lmduan@tsinghua.edu.cn}
\affiliation{Center for Quantum Information, IIIS, Tsinghua University, Beijing 100084, PR China}

\begin{abstract}
We report an experimental demonstration of a machine learning approach to identify exotic topological phases, with a focus on the three-dimensional chiral topological insulators. We show that the convolutional neural networks---a class of deep feed-forward artificial neural networks with widespread applications in machine learning---can be trained to successfully identify different topological phases protected by chiral symmetry from experimental raw data generated with a solid-state quantum simulator.  Our results explicitly showcase the exceptional power of machine learning in the experimental detection of topological phases, which paves a way to study rich topological phenomena with the machine learning toolbox.
\end{abstract}

\date{\today}
\maketitle

The discovery of topological phases has transformed our modern understanding of quantum phases of matter \cite{von1986quantized, Hasan2010Colloquium, Qi2011Topological}. Unlike conventional phases, topological phases do not fit into the paradigm of symmetry breaking \cite{Lifshitz2013Statistical}. Instead, distinct topological phases exist within each symmetry class~\cite{Kitaev2009Periodic, Schnyder2008Classification, Ryu2010Topological, Chiu2016Classification}.
In the experiment, the identification of topological phases typically relies on measuring certain properties that have a topological origin.  
For example, to identify the three-dimensional (3D) $\mathbb Z_{2}$ topological insulators (TIs), one can measure their robust surface states (e.g., Dirac cones) \cite{Hsieh2008topological, Chen2009Experimental}. 
Yet, for other topological phases, it can be technologically challenging to provide a nonambiguous experimental identification.

Machine learning, or more broadly speaking artificial intelligence, may provide new promising approaches to solve such challenges. 
In fact, its ideas and techniques have recently been explored in condensed matter physics, giving rise to a fresh research frontier of machine learning phases of matter \cite{ Carleo2017Solving, Deng2017Machine,Gao2017Efficient,Deng2017Quantum, Wang2016Discovering, Carrasquilla2017Machine, Nieuwenburg2017Learning, Chng2017Machine, Zhang2017Quantum, Zhang2018Machine,ZhangYi2017ML, Broecker2017Machine,Wetzel2017Unsupervised,Hu2017Discovering}. Particularly, theoretical proposals for using supervised learning to identify topological phases have been proposed and have attracted considerable attention across multiple communities \cite{Zhang2017Quantum, Zhang2018Machine, ZhangYi2017ML}. A striking advantage of this approach is that the neural networks can be trained to detect topological phases directly from raw data of local observables, such as spin textures or local correlations, though the topological properties are intrinsically nonlocal. Despite all this exciting theoretical progress, the power of machine learning in studying quantum phases of matter, especially topological phases, has not yet been demonstrated in laboratory.

In this Letter, we carry out such an experiment and demonstrate that neural networks can be trained to successfully identify topological phases,  even with a very small portion of the experimentally generated raw data. By using the electron spins in a diamond nitrogen-vacancy (NV) center, we implement a model Hamiltonian for 3D chiral TIs and obtain the corresponding single-particle density matrices in the momentum space through quantum state tomography~\cite{Yuan2017Observation}. Using the measured density matrices as input data, we find that a trained 3D convolutional neural network (CNN) can correctly identify distinct topological phases. We emphasize that this approach only requires a small portion of data samples, which would significantly reduce the cost in practical experiments \cite{DataNumber}.  Our results demonstrate the experimental feasibility of the machine learning approach to study topological phases, which opens the door for experimental detection of exotic quantum phases with the machine learning toolbox.

We focus our discussion on 3D chiral TIs, which have attracted extensive attention in recent years \cite{Hosur2010Chiral, Neupert2012Noncommutative, Shiozaki2013Electromagnetic, Wang2014Probe, Hasebe2014Chiral, Wang2015Quantized} . Unlike the well-known $\mathbb Z_{2}$ TIs, chiral TIs break both time-reversal and particle-hole symmetry, but preserve chiral symmetry and are characterized by a $\mathbb Z$ topological invariant \cite{Kitaev2009Periodic, Schnyder2008Classification, Ryu2010Topological, Chiu2016Classification}.
Due to chiral symmetry, chiral TIs exhibit a zero-energy flat band when the total number of bands is odd. Strong interactions may lead to nontrivial many-body effects in 3D topological flat bands \cite{Weeks2012Flat, Neupert2012Noncommutative, Wang2014Probe} and chiral TIs could provide an ideal setting to study possible fractionalization in 3D \cite{Maciejko2010Fractional, Swingle2011Correlated, Wang2014Classification}. In our experiment, we detect the topological flat band via energy dispersion measurements. Topological phase transitions and the accompanied gap closings are also observed by changing a control parameter in the Hamiltonian.

We consider a minimal three-band Hamiltonian for 3D chiral TIs $H = \sum_{\mathbf{k} \in \text{BZ}} \Psi_{\mathbf k}^{\dagger} \mathcal H_{\mathbf k} \Psi_{\mathbf k}$, where $\Psi_{\mathbf k}^{\dagger} = (c_{\mathbf k, 1}^{\dagger},c_{\mathbf k, 0}^{\dagger},c_{\mathbf k, -1}^{\dagger})$ with $c_{\mathbf k, \mu}^{\dagger}$ creating a fermion at momentum $\mathbf k = (k_{x}, k_{y}, k_{z})$ in the orbital (spin) state $\mu = 1,0,-1$ and the summation is over the Brillouin zone (BZ); the single-particle momentum-resolved Hamiltonian is \cite{Neupert2012Noncommutative, Wang2014Probe}: 
\begin{equation}
\label{Eq:Ham}
\mathcal H_{\mathbf k} = \left(
\begin{array}{ccc}
0  & q_{1}(\mathbf k) - iq_{2}(\mathbf k) & 0 \\ q_1(\mathbf k) + iq_2(\mathbf k) & 0 & q_3(\mathbf k) + iq_0(\mathbf k) \\ 0 & q_3(\mathbf k)-iq_0(\mathbf k) & 0 \end{array}
\right)
\end{equation}
with momentum-dependent functions $q_{1}(\mathbf k) = t\sin k_{x}$, $q_{2}(\mathbf k) = t\sin k_{y}$, $q_{3}(\mathbf k) = t\sin k_{z}$, and $q_{0}(\mathbf k) = t(\cos k_{x} +\cos k_{y} + \cos k_{z} +h)$ where $t = 1$ is set to be the energy unit and $h$ is a dimensionless control parameter. The Hamiltonian has a chiral symmetry $S H_{\mathbf k} S^{-1} = - H_{\mathbf k}$ specified by the unitary transformation $S \equiv \text{diag}(1,1,-1)$. The topological properties for each band can be characterized by a topological invariant $\theta^{(\lambda)}$  \cite{TopInv}, which has measurable consequences in 3D TIs, such as the magneto-electric polarization \cite{Qi2008Topological, Essin:2009ui, Shiozaki2013Electromagnetic, Wang2015Quantized}. For the middle band, direct calculations show that $\theta^{(m)}/\pi=0,1$ and $-2$ for $|h| >3$, $1<|h|<3$, and $|h| <1$, respectively.

\begin{figure}[t]
\includegraphics[trim=0cm 0cm 0cm 0cm, clip,width=\linewidth]{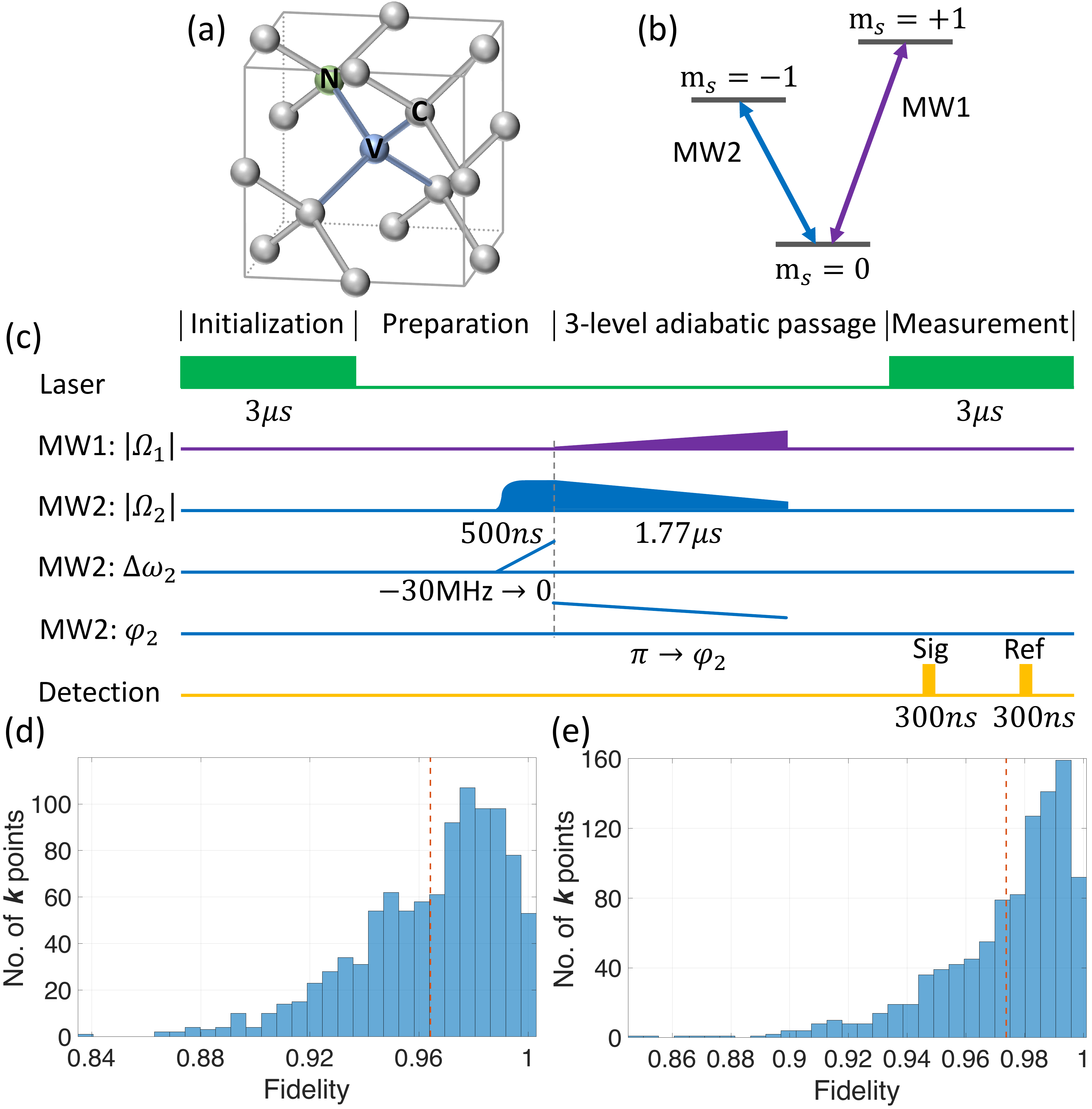}
\caption{(a) Structure of the nitrogen-vacancy (NV) center. (b) Ground state energy levels with their corresponding electron spins, coupled by two microwaves (MW1 and MW2), are used to simulate the three-band topological Hamiltonian. (c) An experimental sequence for the adiabatic passage. The amplitude of MW1, amplitude of MW2, detuning of MW2, and phase of MW2 are denoted as $|\Omega_{1}|$, $|\Omega_{2}|$, $\Delta \omega_{2}$ and $\varphi_{2}$, respectively. The phase of MW1 is kept constant. This path illustrates the adiabatic preparation for the ground state of the Hamiltonian $\mathcal H_{\mathbf k}$ at $(k_{x},k_{y},k_{z}) = (0.6\pi, 0.6\pi, 0.6\pi)$ and $h=1$. The spin is first initialized to $\ket{0}$ and then adiabatically ramped into the ground state of $\mathcal H_{\mathbf k}$ at $\mathbf k = \mathbf 0$ in the preparation stage. Next, the Hamiltonian $\mathcal H_{\mathbf k}$ is slowly ramped from $\mathbf k = \mathbf 0$ to each specific $\mathbf k$ by controlling the amplitude and phase of MW1 and MW2. Lastly, full quantum state tomography is performed through measurements in various bases. (d)-(e) Histogram of the measured state fidelity $F_{\mathbf k}$ at each $\mathbf k$ on a $10 \times 10 \times 10$ regular grid for the middle band with (d) $h = 2$ and (e) $h = 4$. The average fidelity is $96.3(27)\%$ and $97.4(23)\%$ for $h = 2$ and $ h = 4$, respectively, indicated by the dashed lines.}
\label{Fig:FigExp}
\end{figure}

We experimentally realize this model Hamiltonian and probe its topological characteristics using a single NV center in a diamond sample at room temperature [Fig.~\ref{Fig:FigExp}(a)]. Our experimental setup is based on a home-built confocal microscope with an oil-immersed objective lens. The three spin (orbital) degrees of freedom can be encoded into the ground state $\ket{1}, \ket{0}, \ket{-1}$ corresponding to $m_{s} = 1,0, -1$ electron spin states of the NV center [Fig.~\ref{Fig:FigExp}(b)]. A magnetic field is applied along the NV symmetry axis to polarize nearby spins and remove the degeneracy between the $m_{s} = 1$ and $m_{s} = -1$ states. With no inter-particle interactions, the Hamiltonian $H$ is diagonal in the momentum space, so different momentum components in each band are decoupled. One can thus measure each momentum component separately in the experiment and probe the properties of the entire topological band by collecting individual measurements.

From Eq.~\eqref{Eq:Ham}, one observes that the Hamiltonian $\mathcal H_{\mathbf k}$ has no diagonal terms and no direct coupling between states $\ket{1}$ and $\ket{-1}$, so we can implement the Hamiltonian at each $\mathbf k$ by two on-resonance microwaves as shown in Fig.~\ref{Fig:FigExp}(b)~\cite{SM}. In our experiment, we do not have a physical analog for the lattice momentum $\mathbf k$. Instead, an adiabatic passage is used to realize the eigenstate of each band of the Hamiltonian $\mathcal H_{\mathbf k}$ at some parametric point $\mathbf k$. Fig.~\ref{Fig:FigExp}(c) shows such an experimental sequence. The spin state is first initialized to the state $\ket{0}$. To probe the properties of the lower band, we first adiabatically prepare the state $\ket{\Psi^{(l)}_{0}} = \frac{1}{\sqrt{2}} \left(\ket{0} + i \ket{{-1}} \right)$, which is the ground state of the Hamiltonian $\mathcal H_{\mathbf k}$ at $\mathbf k =(0,0,0)$. Then ground states at other  $\mathbf k$ points  can be attained by adiabatically tuning the microwaves to the final Hamiltonian $\mathcal H_{\mathbf k}$~\cite{SM}. For the middle band, the state $\ket{\Psi^{(m)}_{0}} = \ket{1}$ is first prepared, which is the middle eigenstate of the Hamiltonian at $\mathbf k =(0,0,0)$, before tracing a similar adiabatic passage to reach any other states of the flat band. Interestingly, the middle-band states are effectively the dark state of the Hamiltonian $\mathcal H_{\mathbf k}$ at all $\mathbf k$  points, which offers an alternative interpretation for the macroscopic flat band. Thus, the adiabatic procedure to reach any middle-band eigenstates resembles the stimulated Raman adiabatic passage (STIRAP). It is intriguing that the energy band composed of dark states can be topologically nontrivial, as we demonstrate explicitly later by learning the topological invariants and probing of phase transitions. The topological gap between different bands guarantees the feasibility of the above adiabatic procedure, and in order to remain as adiabatic as possible, we ramp the microwaves more slowly when the gap is smaller. This also enables us to probe the system close to topological phase transitions.

After the adiabatic passage, we perform quantum state tomography for the qutrit system via measurements in eight different sets of bases, and retrieve the full density matrix via a maximum likelihood estimation~\cite{SM}. Through these tomography measurements on a momentum grid, we obtain full information about the corresponding band. Figs.~\ref{Fig:FigExp}(d) and \ref{Fig:FigExp}(e) show histograms of the retrieved state fidelity $F_{\mathbf k} = \bracket{\Psi_{\mathbf k}}{\rho_{\mathbf k}}{\Psi_{\mathbf k}}$ for the middle band on a $10 \times 10 \times 10$ regular grid, where $\Psi_{\mathbf k}$ is the theoretically obtained eigenstate and $\rho_{\mathbf k}$ is the experimentally measured density matrix, which serves as the input data for the neural network.

\begin{figure}[t!]
\includegraphics[trim=0cm 0cm 0cm 0cm, clip,width=\linewidth]{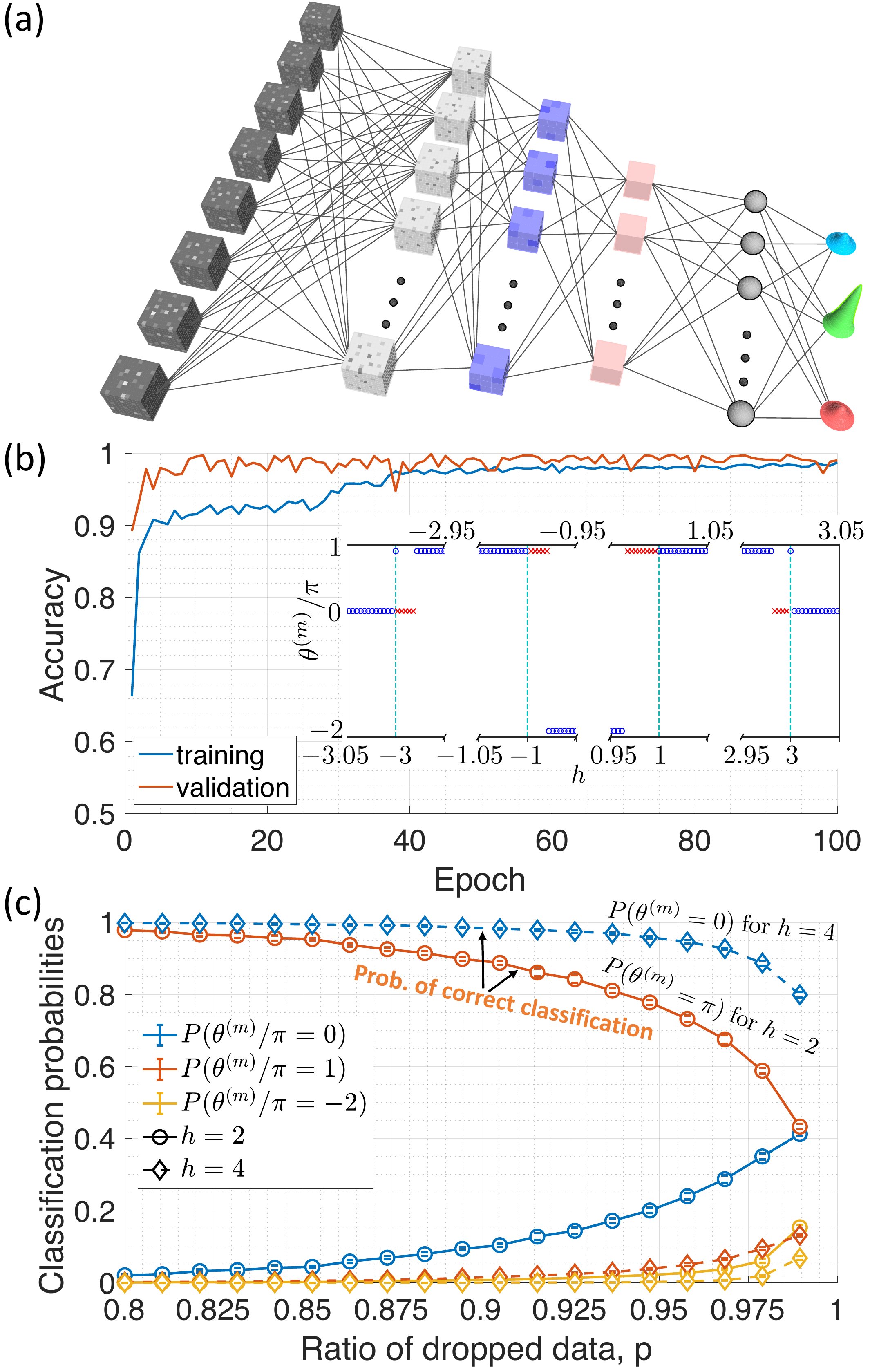}
\caption{Machine learning topological phases. (a) Architecture of the 3D convolutional neural network (CNN) to classify the topological phases. The input is experimental data of density matrices on a $10 \times 10 \times 10$ regular grid. Each density matrix is represented by 8 real numbers~\cite{SM}. The output is the classification probabilities for each possible phase with $\theta^{(m)}/\pi = 0,1,-2$. (b) The CNN is trained with numerically simulated data generated uniformly from $h \in [-5,5]$ while avoiding the intervals $[1.8,2.2]$ and $[3.8,4.2]$ (where experimental data lie in). There are 3680 training samples and 921 validation samples. The inset shows that wrong classifications (red crosses) cluster near the points of topological phase transitions. (c) CNN classification of the experimental data with $h = 2$ and $h=4$. We randomly discard a proportion $p$ out of the 1000 measured density matrices for each $h$ (the missing data are replaced with zero values to keep the input size fixed). Error bars are obtained with 500 random data dropping process.}
\label{Fig:CNN}
\end{figure}

\begin{figure*}[!t]
\includegraphics[trim=0cm 0cm 0cm 0cm, clip,width=\linewidth]{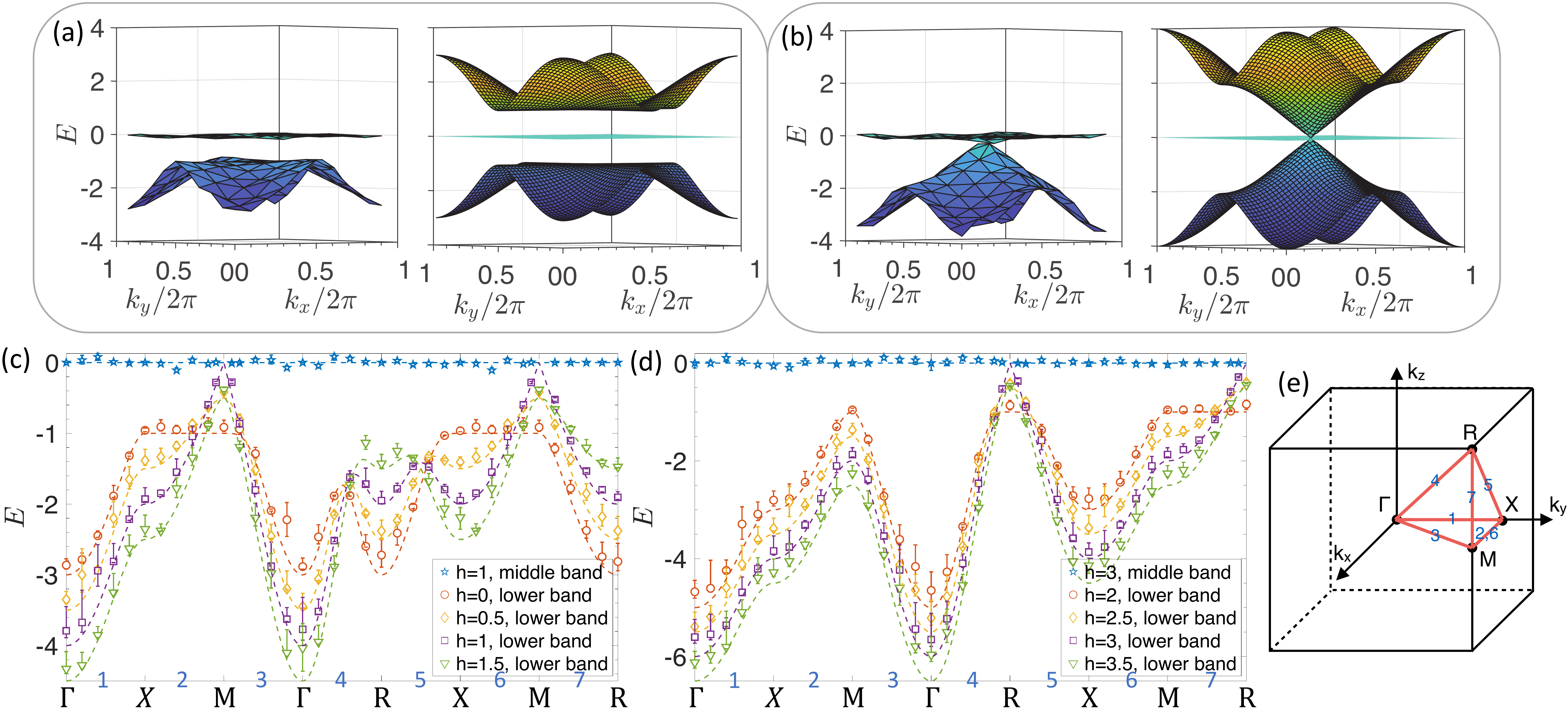}
\caption{Topological flat band and topological phase transitions. (a) Experimentally measured energy dispersion of the middle and lower band for the momentum layer $k_{z}=\pi$ with $ h = 2$ (left panel). Energies are measured on a $10 \times 10$ grid in the $(k_{x},k_{y})$ plane, and the surface plots are interpolated from the grid. Corresponding theoretical energy dispersion is shown in the right panel. (b) same as (a) but for $h = 3$. (c-d) Energy spectrum of the system for different values of the parameter $h$, showing gap-closing topological phase transitions at $h=1,3$. The high symmetry points are respectively $\Gamma = (0,0,0)$, $X = (0,\pi,0)$, $M = (\pi,\pi,0)$, and $R= (\pi,\pi,\pi)$. Markers are experimental data with error bars representing uncertainties obtained from the $95\%$ confidence interval. Dashed lines are theoretical lines. (e) Brillouin zone showing the high symmetry points and the path connecting them.}
\label{Fig:Energy}
\end{figure*}

In the experiment, we use a pre-trained CNN to predict the topological invariant. This is in the setting of supervised learning, where the essential idea is the same as how one trains artificial neural networks to identify pictures of dogs and cats. A CNN is first trained with numerically sampled data (density matrices)  from $h \in [-5,5]$, avoiding the intervals $[1.8,2.2]$ and $[3.8,4.2]$ where experimental data lie. Correct values of the  topological invariant $\theta^{(m)}/\pi$ are numerically computed and assigned to the corresponding training samples. Fig.~\ref{Fig:CNN}(a) shows a schematic of the CNN we used, where the input data are the state density matrices on a $10 \times 10 \times 10$ grid, and the output reports the classification probabilities $P(\theta^{m}/\pi = 0,1,-2)$ for each possible phase~\cite{SM}. Fig.~\ref{Fig:CNN}(b) shows the training and validation accuracies with increasing iterations. Both the training and validation accuracies increase rapidly at the beginning of the training process and then saturate at a high value ($\approx 98\%$). Near the topological phase transition points, wrong classifications may occur due to large quantum fluctuations, as observed in the inset of Fig.~\ref{Fig:CNN}(b). 

After training, we use the CNN to predict the topological phases of our experimental data at $h=2$ and $h=4$. We find that, when using the entire set of measured density matrices on the $10 \times 10 \times 10$ grid, the trained CNN can identify the topological phase with a nearly unity probability. A more interesting scenario arises when part of the data is missing or not accessible. This is inspired by the fact that in the case of  learning to recognize dogs and cats, the trained network can sometimes successfully recognize images that are partially destroyed. As shown in Fig.~\ref{Fig:CNN}(c), it is remarkable that the trained CNN can successfully identify (with a probability $\gtrsim 90\%$) different topological phases even with less than $10\%$ of the experimental data.

We remark that the machine learning approach outshines conventional methods in several aspects. First, the neural network needs not to know the precise formula for different topological invariants. What is necessary is the labelling of the training data. Thus, this approach has a broader applicability in general. Especially, it carries over to the cases of poorly understood phases, where the precise order parameters or topological characterization are not known a priori. Second, in terms of data efficiency, CNN beats conventional methods that use discrete integration, which is not directly applicable if data are not supplied on a regular grid~\cite{SM}. In practice, this would reduce the experimental cost dramatically since much fewer data samples are required. Third, machine learning tools have the advantage of directly using measurement results in the form of mixed states, whereas conventional methods typically rely on topological invariants commonly defined in terms of pure states.  In addition, we mention that CNN can be used to identify topological phase transition points, despite the fact that the training data are sampled far away from the transition boundaries~\cite{SM}.

As mentioned before, an interesting property of this three-band chiral TI is the presence of a three-dimensional macroscopic flat band. To probe this dispersionless band, we measure the energy of the middle band via state tomography. Left panels of Fig.~\ref{Fig:Energy}(a) and (b) show the measured energy spectrum on a $10 \times 10$ grid in the $(k_{x},k_{y})$ plane for the momentum layer $k_{z} = \pi$. For other layers of $k_{z}$, we also observe the flat middle band, confirming the 3D nature of the macroscopic flat band. For $h = 2$, we can clearly see a band gap, while for $h = 3$, the gap closes at $(k_{x},k_{y},k_{z}) = (\pi, \pi, \pi)$, signaling the topological phase transition. 

To further probe these topological phase transitions, we measure the energies of the lower band and the middle band along a path connecting the high symmetry points in the momentum space for various values of the parameter $h$. Fig.~\ref{Fig:Energy}(c) and (d) show the measured energy spectrum along the smooth path $\Gamma$-$X$-$M$-$\Gamma$-$R$-$X$-$M$-$R$, where $\Gamma,X,M,R$ are high-symmetry momentum points as illustrated in Fig.~\ref{Fig:Energy}(e). By changing the parameter from $h=0$ to $h=3.5$, we see the gap between the lower band and the middle band closes at the $M$ point for $h = 1$ [Fig.~\ref{Fig:Energy}(c)] and at the $R$ point for $h = 3$  [Fig.~\ref{Fig:Energy}(d)], consistent with our theoretical predictions. The gapless points are deliberately avoided due to the breakdown of adiabaticity, but the trend of gap closing can be clearly seen. The topological phase transition points can also be successfully predicted by the trained CNN~\cite{SM}.

In conclusion, we have experimentally demonstrated a machine learning approach to identify topological phases with a solid-state quantum simulator. We show that a CNN can be trained to successfully identify topological phases, even with a minimal number of data samples.  This demonstrates a striking advantage of the machine learning approach, which could significantly reduce the experimental cost. Since the CNN approach does not require knowing the formula for topological invariants, it generalizes directly to other exotic phases that are less well-understood.

Although we have explicitly focused on interaction-free topological phases, extensions including the effect of interaction are possible. One approach is to embed the interaction-free Hamiltonian into the parameter space of multiple qutrits and introduce interactive couplings between the qutrits, similar to the method in Ref.~\cite{Roushan2014Observation}. This represents an exciting possibility to probe the effect of interactions in 3D topological flat bands in such an interacting spin system. The use of modern machine learning tools in this setting may offer new insights in understanding topological phases with strong interactions.

\begin{acknowledgments}
This work was supported by the National key Research and Development Program of China (2016YFA0301902), Tsinghua University, and the Ministry of Education of China.
\end{acknowledgments}


\begin{thebibliography}{99}
  \bibitem{von1986quantized}
K. von Klitzing,
 \href{https://link.aps.org/doi/10.1103/PhysRevB.96.245119}{Rev. Mod. Phys. \textbf{58,} 519 (1986).}
\bibitem{Hasan2010Colloquium}
M. Z. Hasan and C. L. Kane,
 \href{http://link.aps.org/doi/10.1103/RevModPhys.82.3045}{Rev. Mod. Phys. \textbf{82,} 3045 (2010).}
\bibitem{Qi2011Topological}
X.-L. Qi and S.-C. Zhang,
 \href{http://link.aps.org/doi/10.1103/RevModPhys.83.1057}{Rev. Mod. Phys. \textbf{83,} 1057 (2011).}
\bibitem{Lifshitz2013Statistical}
E. M. Lifshitz and L. P. Pitaevskii, \textit{Statistical physics: theory of the condensed state} (Butterworth-Heinemann, Oxford, 2013), Vol. 9.
\bibitem{Kitaev2009Periodic}
A. Kitaev,
 \href{http://scitation.aip.org/content/aip/proceeding/aipcp/10.1063/1.3149495}{AIP Conf. Proc. \textbf{1134,} 22 (2009).}
\bibitem{Schnyder2008Classification}
A. P. Schnyder, S. Ryu, A. Furusaki, and A. W. W. Ludwig,
 \href{http://link.aps.org/doi/10.1103/PhysRevB.78.195125}{Phys. Rev. B \textbf{78,} 195125 (2008).}
\bibitem{Ryu2010Topological}
S. Ryu, A. P. Schnyder, A. Furusaki, and A. W. W. Ludwig,
 \href{http://stacks.iop.org/1367-2630/12/i=6/a=065010}{New J. Phys. \textbf{12,} 065010 (2010).}
 \bibitem{Chiu2016Classification}
C.-K. Chiu, J. C. Y. Teo, A. P. Schnyder, and S. Ryu,
 \href{https://link.aps.org/doi/10.1103/RevModPhys.88.035005}{Rev. Mod. Phys. \textbf{88,} 035005 (2016).}
  \bibitem{Hsieh2008topological}
D. Hsieh, D. Qian, L. Wray, Y. Xia, Y. S. Hor, R. J. Cava, and M. Z. Hasan,
 \href{http://dx.doi.org/10.1038/nature06843}{Nature \textbf{452,} 970 (2008).}
  \bibitem{Chen2009Experimental}
Y. L. Chen, J. G. Analytis, J.-H. Chu, Z. K. Liu, S.-K. Mo, X. L. Qi, H. J. Zhang, D. H. Lu, X. Dai, Z. Fang, S. C. Zhang,
I. R. Fisher, Z. Hussain, and Z.-X. Shen,
 \href{http://science.sciencemag.org/content/325/5937/178}{Science \textbf{325,} 178 (2009).}
  \bibitem{Carleo2017Solving}
G. Carleo and M. Troyer,
 \href{http://science.sciencemag.org/content/355/6325/602}{Science \textbf{355,} 602 (2017).}
   \bibitem{Deng2017Machine}
D.-L. Deng, X. Li, and S. Das Sarma,
 \href{https://link.aps.org/doi/10.1103/PhysRevB.96.195145}{Phys. Rev. B \textbf{96,} 195145 (2017).}
 \bibitem{Gao2017Efficient}
X. Gao and L. M. Duan,
 \href{https://doi.org/10.1038/s41467-017-00705-2}{Nat. Commun. \textbf{8,} 662 (2017).}
  \bibitem{Deng2017Quantum}
D.-L. Deng, X. Li, and S. Das Sarma,
 \href{https://link.aps.org/doi/10.1103/PhysRevX.7.021021}{Phys. Rev. X \textbf{7,} 021021 (2017).}
 \bibitem{Wang2016Discovering}
L. Wang,
 \href{https://link.aps.org/doi/10.1103/PhysRevB.94.195105}{Phys. Rev. B \textbf{94,} 195105 (2016).}
 \bibitem{Carrasquilla2017Machine}
J. Carrasquilla and R. G. Melko,
 \href{http://dx.doi.org/10.1038/nphys4035}{Nat. Phys. \textbf{13,} 431 (2017).}
 \bibitem{Nieuwenburg2017Learning}
E. P. L. van Nieuwenburg, Y.-H. Liu, and S. D. Huber,
 \href{http://dx.doi.org/10.1038/nphys4037}{Nat. Phys. \textbf{13,} 435 (2017).}
 \bibitem{Chng2017Machine}
K. Ch'ng, J. Carrasquilla, R. G. Melko, and E. Khatami,
 \href{https://link.aps.org/doi/10.1103/PhysRevX.7.031038}{Phys. Rev. X \textbf{7,} 031038 (2017).}
 \bibitem{Zhang2017Quantum}
Y. Zhang and E.-A. Kim,
 \href{https://link.aps.org/doi/10.1103/PhysRevLett.118.216401}{Phys. Rev. Lett. \textbf{118,} 216401 (2017).}
 \bibitem{Zhang2018Machine}
P. Zhang, H. Shen, and H. Zhai,
 \href{https://link.aps.org/doi/10.1103/PhysRevLett.120.066401}{Phys. Rev. Lett. \textbf{120,} 066401 (2018).}
 \bibitem{ZhangYi2017ML}
Y. Zhang, R. G. Melko, and E.-A. Kim,
 \href{https://link.aps.org/doi/10.1103/PhysRevB.96.245119}{Phys. Rev. B \textbf{96,} 245119 (2017).}
 \bibitem{Broecker2017Machine}
P. Broecker, J. Carrasquilla, R. G. Melko, and S. Trebst,
 \href{http://www.nature.com/articles/s41598-017-09098-0}{Sci. Rep. \textbf{7,} 8823 (2017).}
 \bibitem{Wetzel2017Unsupervised}
S. J. Wetzel,
 \href{https://link.aps.org/doi/10.1103/PhysRevE.96.022140}{Phys. Rev. E \textbf{96,} 022140 (2017).}
 \bibitem{Hu2017Discovering}
W. Hu, R. R. P. Singh, and R. T. Scalettar,
 \href{https://link.aps.org/doi/10.1103/PhysRevE.95.062122}{Phys. Rev. E \textbf{95,} 062122 (2017).}
 \bibitem{Yuan2017Observation}
X.-X. Yuan, L. He, S.-T. Wang, D.-L. Deng, F. Wang, W.-Q. Lian, X. Wang, C.-H. Zhang, H.-L. Zhang, X.-Y. Chang,
and L.-M. Duan,
 \href{http://cpl.iphy.ac.cn/EN/abstract/article_70778.shtml}{Chin. Phys. Lett. \textbf{34,} 060302 (2017).}
 
\bibitem{DataNumber}
The number of data points required depends on the phase region that the CNN is attempting to learn. Near phase transitions, more data points are needed due to stronger quantum fluctuations. Quantitative comparisons can be found in Fig.~2c and Fig.~S2. In practice, the data size required is a priori unknown and one needs to increase the number of measurement points until convergence is observed.
 
\bibitem{Hosur2010Chiral}
P. Hosur, S. Ryu, and A. Vishwanath,
 \href{https://link.aps.org/doi/10.1103/PhysRevB.81.045120}{Phys. Rev. B \textbf{81,} 045120 (2010).}
 \bibitem{Neupert2012Noncommutative}
T. Neupert, L. Santos, S. Ryu, C. Chamon, and C. Mudry,
 \href{http://link.aps.org/doi/10.1103/PhysRevB.86.035125}{Phys. Rev. B \textbf{86,} 035125 (2012).}
 \bibitem{Shiozaki2013Electromagnetic}
K. Shiozaki and S. Fujimoto,
 \href{https://link.aps.org/doi/10.1103/PhysRevLett.110.076804}{Phys. Rev. Lett. \textbf{110,} 076804 (2013).}
  \bibitem{Wang2014Probe}
S.-T. Wang, D.-L. Deng, and L.-M. Duan,
 \href{http://link.aps.org/doi/10.1103/PhysRevLett.113.033002}{Phys. Rev. Lett. \textbf{113,} 033002 (2014).}

\bibitem{Hasebe2014Chiral}
K. Hasebe,
 \href{http://www.sciencedirect.com/science/article/pii/S0550321314002247}{Nucl. Phys. B \textbf{886,} 681 (2014).}
 \bibitem{Wang2015Quantized}
S.-T. Wang, D.-L. Deng, J. E. Moore, K. Sun, and L.-M. Duan,
 \href{http://link.aps.org/doi/10.1103/PhysRevB.91.035108}{Phys. Rev. B \textbf{91,} 035108 (2015).}
 
 \bibitem{Weeks2012Flat}
C. Weeks and M. Franz,
 \href{https://link.aps.org/doi/10.1103/PhysRevB.85.041104}{Phys. Rev. B \textbf{85,} 041104 (2012).}
  \bibitem{Maciejko2010Fractional}
J. Maciejko, X.-L. Qi, A. Karch, and S.-C. Zhang,
 \href{https://link.aps.org/doi/10.1103/PhysRevLett.105.246809}{Phys. Rev. Lett. \textbf{105,} 246809 (2010).}
  \bibitem{Swingle2011Correlated}
B. Swingle, M. Barkeshli, J. McGreevy, and T. Senthil,
 \href{https://link.aps.org/doi/10.1103/PhysRevB.83.195139}{Phys. Rev. B \textbf{83,} 195139 (2011).}
\bibitem{Wang2014Classification}
C. Wang, A. C. Potter, and T. Senthil,
 \href{http://science.sciencemag.org/content/343/6171/629}{Science \textbf{343,} 629 (2014).}




\bibitem{TopInv}
For each band, a topological invariant $\theta^{(\lambda)}$ can characterize the topological attribute of the Bloch states, where $\lambda = l,m,u$ denotes the lower, middle and upper bands, respectively. The invariant can be written as an integral in the three-dimensional momentum space
\begin{equation}
\theta^{(\lambda)} = \dfrac{1}{4\pi} \int_{ \text{BZ} }  \epsilon^{\mu \nu \tau } A_{\mu}^{(\lambda)} \partial_{k^{\nu}} A_{\tau}^{(\lambda)}  \, d^{3}k   \label{TopInvar}
\end{equation}
where $\epsilon^{\mu \nu \tau }$ is the Levi-Civita symbol with $\mu ,\nu ,\tau \in \left\{ x,y,z\right\}$, and the Berry connection is $A_{\mu}^{(\lambda)} = \bra{\psi_{\mathbf k}^{(\lambda)}} \partial_{k^{\mu}} \ket{\psi_{\mathbf k}^{(\lambda)}} $ where $\ket{\Psi _{\mathbf{k}}^{(\lambda)}}$ denotes the Bloch state for the $\lambda$ band. The topological invariants for each band are related as $\theta^{(u)} = \theta^{(l)} = \theta^{(m)}/4$ for the studied three-band model \cite{Neupert2012Noncommutative}.



\bibitem{Qi2008Topological}
X.-L. Qi, T. L. Hughes, and S.-C. Zhang,
 \href{https://link.aps.org/doi/10.1103/PhysRevB.78.195424}{Phys. Rev. B \textbf{78,} 195424 (2008).}
 \bibitem{Essin:2009ui}
A. M. Essin, J. E. Moore, and D. Vanderbilt,
 \href{http://link.aps.org/doi/10.1103/PhysRevLett.102.146805}{Phys. Rev. Lett. \textbf{102,} 146805 (2009).}
 

 \bibitem{SM}
See Supplemental Material for more details on the adiabatic procedure, qutrit state tomography, description of the CNN, and additional numerical simulations.
  
 \bibitem{Roushan2014Observation}
P. Roushan, C. Neill, Y. Chen, M. Kolodrubetz, C. Quintana, N. Leung, M. Fang, R. Barends, B. Campbell, Z. Chen,
B. Chiaro, A. Dunsworth, E. Jeffrey, J. Kelly, A. Megrant, J. Mutus, P. J. J. O'Malley, D. Sank, A. Vainsencher, J. Wenner,
T. White, A. Polkovnikov, A. N. Cleland, and J. M. Martinis,
 \href{http://dx.doi.org/10.1038/nature13891}{Nature \textbf{515,} 241 (2014).}

\end{thebibliography}
\end{document}


\title{Supplemental Material: Machine learning topological phases with a solid-state quantum simulator}

\author{Wenqian Lian}\thanks{These authors contributed equally to this work.}
\affiliation{Center for Quantum Information, IIIS, Tsinghua University, Beijing 100084, PR China}
\author{Sheng-Tao Wang}\thanks{These authors contributed equally to this work.}
\affiliation{Department of Physics, Harvard University, Cambridge, MA 02138, USA}
\affiliation{Center for Quantum Information, IIIS, Tsinghua University, Beijing 100084, PR China}
\author{Sirui Lu}\author{Yuanyuan Huang}\author{Fei Wang}\author{Xinxing Yuan}\author{Wengang Zhang}\author{Xiaolong Ouyang}\author{Xin Wang}\author{Xianzhi Huang}\author{Li He}\author{Xiuying Chang}
\author{Dong-Ling Deng}\email{dldeng@mail.tsinghua.edu.cn}
\author{Luming Duan}\email{lmduan@tsinghua.edu.cn}
\affiliation{Center for Quantum Information, IIIS, Tsinghua University, Beijing 100084, PR China}

\date{\today}
\maketitle

\section{Experimental Setup}

Our setup is based on a home-built confocal microscope with an oil-immersed objective lens with numerical aperture $1.49$. A $532\,$nm diode laser, controlled by an acoustic optical modulator (AOM) under the double pass configuration, is used for spin state initialization and detection. The fluorescence photons are coupled into a single-mode fibre and detected with a single photon detector.

Two signal generators (Keysight N5181B) are used to generate microwaves with different frequencies.
To control the amplitude and relative phase of signals, each microwave is modulated by an IQ mixer (Marki IQ 1545LMP) with two orthogonal Megahertz microwaves, which are generated by a power splitter from the same analog output of an arbitrary waveform generator (AWG, Tektronix AWG70002A). The generated microwave signals are amplified by their respective high power amplifiers and subsequently merged into one transmission line by a combiner. The combined signal is then delivered into an impedance matching gold coplanar waveguide deposited on a coverslip where the sample is mounted. Two signal generators and AWG are synchronized by a $10\,$MHz reference rubidium clock.

The sample is a [100]-oriented type \uppercase\expandafter{\romannumeral2}a single crystal diamond grown by chemical vapor deposition (Element 6). It is irradiated by $10\,$MeV electron beam with a dosage of $8\times10^{13} \, e/cm^{2}$ and annealed at 850$^{\circ}$C for 2 hours in high vacuum. To enhance the collection efficiency, a solid immersion lens with 5.2$\,\mu$m diameter was fabricated on top of the NV center by a focused ion beam. Under $0.35\,$mW laser excitation, the NV center used in this experiment has a flourescence count $\sim 350\,$kcps. A magnetic field of $B_z\approx497.5\,$G is applied along the NV symmetry axis by a permanent magnet to polarize nearby spins~\cite{Smeltzer2009Robust} and remove the degeneracy between the $m_s=+1$ and $m_s=-1$ electron spin states.

\section{Adiabatic passage}

In the experiment, the electronic ground states of the NV center, $\ket{1}, \ket{0}$ and $\ket{-1}$, are used as a qutrit system to simulate the three-band Hamiltonian of chiral topological insulators as written in Eq.~(1) of the main text. The momentum space Hamiltonian is then mapped to the parameter space of our qutrit simulator.

Since the Hamiltonian $\mathcal H_{\mathbf k}$ in Eq.~(1) has no diagonal terms and no direct coupling between states $\ket{1}$ and $\ket{-1}$, we can implement the Hamiltonian at each parametric momentum point by two on-resonance microwaves (MW1 and MW2), coupling states $\ket{1}\leftrightarrow\ket{0}$ and $\ket{-1} \leftrightarrow \ket{0}$, respectively. In the presence of two microwaves and a high magnetic field along $z$ axis, the Hamiltonian for the NV three-level system is
\begin{equation}
H=\Delta S_z^2 + D S_z + |\Omega_1| \sin(\omega_1 t+\phi_1) S_{x1} + |\Omega_2| \sin(\omega_2 t+\phi_2) S_{x2},
\end{equation}
where $\Delta/2\pi = 2.87\,$GHz is the zero-field splitting, $D = g_{NV}\mu_B B_z$ and $S_{z}$ is the spin-1 matrix. $|\Omega_{i}|$, $\omega_i$ and $\phi_{i}$ ($i=1, 2$) denote microwave amplitude, frequency and phase. The subscripts 1 and 2 correspond to two microwaves MW1 and MW2. The spin matrices $S_{x1}$ and $S_{x2}$ correspond to the Pauli-$x$ matrix in the subspace of $\ket{1}, \ket{0}$ and $\ket{0}, \ket{-1}$ respectively, i.e.,
\begin{equation}
S_{x1} = \left(\begin{array}{ccc}0 & 1 & 0 \\1 & 0 & 0 \\0 & 0 & 0\end{array}\right),
\quad
S_{x2} = \left(\begin{array}{ccc}0 & 0 & 0 \\0 & 0 & 1 \\0 & 1 & 0\end{array}\right).
\end{equation}
In the rotating frame of the on-resonance microwaves and applying the rotating wave approximation, the Hamiltonian becomes
\begin{equation}
\begin{aligned}
H_{\text{int}} &= \frac{1}{2}\left( \begin{array}{ccc}
0 & i|\Omega_1| e^{-i \phi_1} & 0 \\
-i|\Omega_1|e^{i \phi_1} & 0 & -i|\Omega_2| e^{i \phi_2} \\
0 & i|\Omega_2| e^{-i \phi_2} & 0
\end{array}
\right) \\
& = \frac{|\Omega_1|}{2}(-\cos\phi_1 S_{y1} +\sin\phi_1 S_{x1})+\frac{|\Omega_2|}{2}(-\cos\phi_2 S_{y2}+\sin\phi_2 S_{x2}),
\end{aligned}
\end{equation}

where $S_{y1}$ and $S_{y2}$ are the corresponding Pauli-$y$ matrix in the subspace of $\ket{1}, \ket{0}$ and $\ket{0}, \ket{-1}$. Comparing with the Hamiltonian for chiral topological insulators [$\mathcal H_{\mathbf k}$ in Eq.~(1)], we see $ q_{1}(\mathbf k) - iq_{2}(\mathbf k) =
\frac{i}{2}|\Omega_1| e^{-i \phi_1}$ and $q_3(\mathbf k) - iq_0(\mathbf k) = \frac{i}{2}|\Omega_2| e^{-i \phi_2}$. Therefore, one can control the amplitude and phase of MW1 and MW2 to realize the topological Hamiltonian at any parametric momentum point $\mathbf k$. The parameter $h$ in Eq.~(1) of the main text is controlled by $|\Omega_2|$ and $\phi_2$.

To simulate the property of a whole band, we prepare the eigenstate of the Hamiltonian at different momentum points via an adiabatic passage. In the experiment, we probe the lower band (ground state band) and middle flat band of the chiral topological Hamiltonian. Before any adiabatic passage, the spin state is first initialized to the state $\ket{0}$. We then prepare the corresponding eigenstate of the Hamiltonian at $k_{x} = k_{y} = k_{z} = 0$, which is $\ket{\psi^{(m)}_0} = \ket{1}$ for the middle band and $\ket{\psi^{(l)}_0} = \frac{1}{\sqrt{2}} (\ket{0} + i\ket{-1})$ for the lower band. We use a $\pi$-pulse between $\ket{1}\leftrightarrow\ket{0}$ state to prepare the initial state $\ket{\psi^{(m)}_0}$. For the lower band, we prepare $\ket{\psi^{(l)}_0}$ by performing an adiabatic passage with the WURST-N ($N=20$) pulse shape \cite{Kupce1995Adiabatic} between $\ket{-1}\leftrightarrow\ket{0}$ state
\begin{align}
|\Omega_2|(t)&= |\Omega_2|_{\max}\left[1-\left|\sin\left( \frac{\pi(T-t)}{2T} \right)\right|^{20}\right], \\
\Delta\omega_2(t) &= \Delta\omega_{2\max} (t/T-1).
\end{align}
Here $T$ is the total sequence time for the two-level adiabatic passage, $|\Omega_2|_{\max} = 2\pi\times 20.83\,$MHz is the maximum amplitude for MW2, and $\Delta\omega_{2\max} = 2\pi\times30\,$MHz is the maximum detuning for the microwave frequency of MW2. Adiabatic condition requires
$\left|\frac{\hbar\braket{\psi^{(m)}}{{\dot{\psi}^{(l)}}}}{E_l-E_m}\right|\ll1$, which means
\begin{equation}
Q\equiv\frac{2(\Delta\omega_2^2+|\Omega_2|^2)^{3/2}}{\dot{|\Omega_2|}\Delta\omega_2-|\Omega_2|\dot{\Delta\omega_2}}\gg1.
\end{equation}
In the preparation of $\ket{\psi^{(l)}_0}$, we have $Q>90$.

After preparing the corresponding initial state for the lower or middle band, we adiabatically change the Hamiltonian from $ \mathbf k = (0,0,0)$ to any other momentum points. By measuring the eigenstate at all momentum points (on a discrete grid) via full quantum state tomography, we can collect individual measurements and hence probe the properties of the entire topological band. More specifically, we first linearly ramp the amplitude of MW2 from $|\Omega_2|_{\max}$ down to the required $|\Omega_2|$, which is normalized by $|\Omega_2|/|\Omega_2|_{\max}=|q_3-iq_0|/|q_3-iq_0|_{\max}$, while the phase changes from $\pi$ to $\phi_2$ for the lower band or keeps constant at $\phi_{2}$ for the middle band. At the same time, the amplitude of MW1, $|\Omega_1|$, is slowly ramped up from 0 to $|\Omega_1|$, determined by $|\Omega_1|/|\Omega_2|=|q_1-iq_2|/|q_3-iq_0|$ with a constant phase $\phi_1$.
The adiabatic condition for our three-level ramping passage is satisfied if
\begin{equation}
Q\equiv\left|\frac{\sqrt{2}(|\Omega_1|^2+|\Omega_2|^2)^{3/2}}{(-|\Omega_2|\dot{|\Omega_1|}+|\Omega_1|\dot{|\Omega_2|}-i|\Omega_1||\Omega_2|\dot{\phi_2})e^{-i(\phi_1+\phi_2)}}\right|\gg1.
\end{equation}
In our experiment, we set $Q>50$.

\section{Quantum state tomography of a qutrit system}

At the end of the adiabatic ramp, we perform a full quantum state tomography to measure the eigenstate of the topological Hamiltonian at some momentum point $\mathbf k$. The density matrix of a qutrit system can be written as
\begin{equation}
\rho = \left[ \begin{array}{ccc}
a & c+id & g+ih\\
c-id & b & e+if \\
g-ih & e-if & 1-a-b
\end{array}
\right].
\end{equation}
The parameters of the density matrix can be determined by measurements in eight different sets of basis as outlined in the table below.
\begin{table}[!hbp]
\centering
\caption{Operations for tomographic measurements}
\begin{tabular*}{10cm}{ccccc}
\toprule
$|1\rangle\leftrightarrow|0\rangle$ & $|-1\rangle\leftrightarrow|0\rangle$ & $p_1$ & $p_0$ & $p_{-1}$\\
\hline
&& $a$ & $b$ & $1-a-b$ \\
$\pi_Y$  &  & $b$ & $a$ & $1-a-b$ \\
$\pi_X/2$  &  & $\frac{a+b}{2}+c$ & $\frac{a+b}{2}-c$ & $1-a-b$ \\
$\pi_Y/2$  &  & $\frac{a+b}{2}-d$ & $\frac{a+b}{2}+d$ & $1-a-b$ \\
  & $\pi_X/2$ & $a$ & $\frac{1-a}{2}-e$ & $\frac{1-a}{2}+e$ \\
  & $\pi_Y/2$ & $a$ & $\frac{1-a}{2}-f$ & $\frac{1-a}{2}+f$ \\
  $\pi_Y$  & $\pi_Y/2$ & $b$ & $\frac{1-b}{2}+g$ & $\frac{1-b}{2}-g$ \\
 $\pi_Y$ & $\pi_X/2$ & $b$ & $\frac{1-b}{2}-h$ & $\frac{1-b}{2}+h$  \vspace{.1cm}\\
\hline \hline
\end{tabular*}
\end{table}

The number $p_{1}, p_0$ and $p_{-1}$ are the probabilities of measuring the $m_s = 1, 0, -1$ states after the corresponding rotations. The total fluorescence $N$ we collect in experiment is
\begin{equation}
N=N_{1}p_{1}+N_0p_0+N_{-1}p_{-1}
\end{equation}
where $N_{1}, N_0$ and $N_{-1}$ are the fluorescence of states $\ket{1}, \ket{0}$ and $\ket{-1}$, which are determined by two Rabi oscillations. Combining measurements in all eight bases, the full density matrix can be reconstructed.

To ensure the density matrix satisfy the positivity condition, the maximum likelihood estimation is employed. Define $\hat{\rho}_p(t)$ as a legal density matrix
\begin{equation}
\hat{\rho}_p(t) = \hat{T}^{\dag}(t)\hat{T}(t)/\text{Tr}[\hat{T}^{\dag}(t)\hat{T}(t)],
\end{equation}
where $\hat{T}$ is a tridiagonal matrix formed by a set of independent real parameters $\{t_1,t_2,\cdots\!,t_9\}$:
\begin{equation}
\hat{T}(t) = \left[ \begin{array}{ccc}
t_1 & 0 & 0\\
t_4+it_5 & t_2 & 0 \\
t_8+it_9 & t_6+it_7 & t_3
\end{array}
\right].
\end{equation}

The measurement data after the above operations are noted as $n_{\nu}$ ($\nu=1,2,\cdots\!,8$) whose expected value is $\bar{n}_{\nu}$. Assume that the noise of measurement data has a Gaussian probability distribution \cite{James2001Measurement}. The probability of obtaining a set of eight counts $\{n_1,n_2,\dots,n_8\}$ is
\begin{equation}
P(n_1,n_2,\dots,n_8)=\prod^8_{\nu=1}\exp\left[-\frac{(n_{\nu}-\bar{n}_{\nu})^2}{2\sigma_{\nu}^2}\right],
\end{equation}
where $\sigma_{\nu}$ is the standard deviation for the $\nu$th measurement data approximated to be $\sqrt{\bar{n}_{\nu}}$ and $\bar{n}_{\nu}$ is the expected fluorescence for $\hat{\rho}_p(t)$ after eight operations:
\begin{equation}
\bar{n}_{\nu}(t_1,t_2,\cdots \!,t_9)=N_{1}p_{1,\nu}+N_0p_{0,\nu}+N_{-1}p_{-1,\nu}.
\end{equation}
Instead of searching for the maximum value of $P(n_1,n_2,\dots,n_8)$, it is more convenient to find the minimum of its negative logarithm:
\begin{equation}
\ell(t_1,t_2,\cdots \!,t_9)=\sum_{\nu=1}^8\frac{[\bar{n}_{\nu}(t_1,t_2,\cdots,t_9)-n_{\nu}]^2}{2\bar{n}_{\nu}(t_1,t_2,\cdots \! ,t_9)}.
\end{equation}
With standard numerical optimization tools, we find the optimum set of $\{t_1,t_2,\cdots\!,t_9\}$ to minimize $\ell(t_1,t_2,\cdots \! ,t_9)$. The corresponding $\hat{\rho}_p$ is the best estimation of the density matrix.

\begin{figure}[t!]
\begin{center}
\includegraphics[trim=0cm 0cm 0cm 0cm, clip,width=12cm]{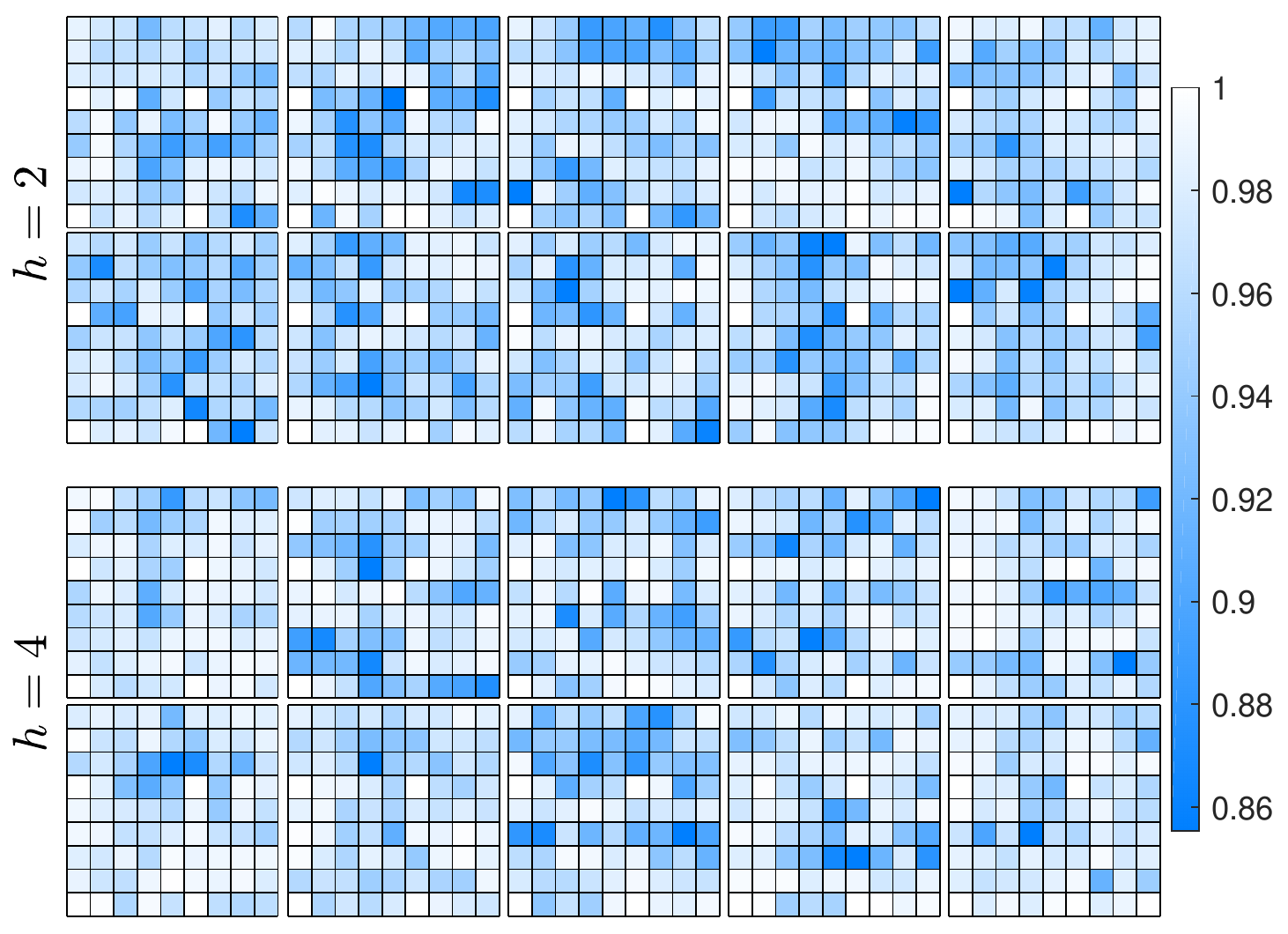}
\caption{State fidelities for the middle band at different momenta $\mathbf k$. The upper and the lower panels have different parameters $h=2$ and $h=4$, respectively. Each panel contains 10 subfigures where the momentum $k_{z}$ are equally spaced from $0$ to $0.9/2\pi$. The horizontal and vertical axes of each subfigure denote $k_{x}$ and $k_{y}$, which vary from 0 to $0.9/2\pi$ with an equal spacing. The average fidelity for the $10\times10\times10$ measured momentum points is $96.3(27)\%$ ($97.4(23)\%$) for the case of $h=2$ ($h=4$).}
\label{Fig:fidelity}
\end{center}
\end{figure}

Each measurement cycle consists of a $3\,\mu$s green laser initialization, microwave operations and final detection with a green laser turned on. During the detection, fluorescence collected in the first $300\,$ns synchronous with the green laser is regarded as the signal, while another $300\,$ns fluorescence collection after $2\,\mu$s laser illumination is the reference. For each measurement, we repeat the experimental cycle for $4\times10^5$ times, collecting a total photon count of $\sim 4\times 10^4$. Error bars are calculated through Monte Carlo simulation by assuming a Poissonian distribution of the counts.

State fidelity is calculated as $F_{\mathbf k} = \bracket{\Psi_{\mathbf k}}{\rho_{\mathbf k}}{\Psi_{\mathbf k}}$, where $\rho_{\mathbf k}$ is the measured state and $\Psi_{\mathbf k}$ is the theoretically obtained state. In Fig.~1(d) and (e) of the main text, we show the histogram of the fidelity distribution. Here, we plot the full state fidelity information at all momenta $\mathbf k$ in Fig.~\ref{Fig:fidelity}. The average fidelity for the $10\times10\times10$ measured momentum points is $96.3(27)\%$ ($97.4(23)\%$) for the case of $h=2$ ($h=4$). The largest contributions for the infidelity are: microwave pulse errors caused by nonlinear equipments such as the combiner and power splitters; environmental fluctuation during the long time measurement as the $N_{1}, N_0$ and $N_{-1}$ are calibrated every $4\sim5$ hours.

\section{Convolutional Neural Network}
\subsection{Descriptions}

As an input to the convolutional neural network (CNN)~\cite{krizhevsky2012imagenet}, we represent a \(n\)-by-\(n\) density matrix \(\rho\) as a real vector \(\mathbf{x}\) in \(\mathbb{R}^{n^2-1}\). In our case, we have a qutrit system where $n = 3$. The density matrix can be expanded in terms of generalized Gell-Mann matrices \cite{bertlmann2008bloch} and the identity. Let us briefly recall the definition of generalized Gell-Mann matrix here.

Let \(\left\{ |{1}\rangle, \ldots, |{n}\rangle \right\}\) be the
computational basis of the \(n\)-dimensional Hilbert space, and
\(E_{j,k} = |{j}\rangle\langle{k}|\). We now define three sets of matrices: the first set is symmetric, $s_{j,k} = E_{j,k} + E_{k,j}$ for \(1 \le j < k \le n\); the second set is antisymmetric, $a_{j,k} = -i\left(E_{j,k} - E_{k,j}\right)$ for \(1 \le j < k \le n\); the last set is diagonal, $d_{l} = \sqrt{\frac{2}{l(l+1)}} \left( \sum_{j=1}^{l} E_{j,j} - l E_{l+1,l+1} \right)$ for \(1 \le l \le n-1\). The generalized Gell-Mann matrices are elements in the set \(\left\{\lambda_i\right\} = \left\{s_{j,k}\right\} \cup \left\{a_{j,k}\right\} \cup \left\{ d_l \right\}\), which gives a total of \(n^2-1\) matrices.

\(\left\{\lambda_i\right\} \cup \left\{\mathbb{I}\right\}\)
forms an orthogonal basis of observables in the \(n\)-dimensional Hilbert space because $\text{tr}\left(\lambda_i\right) = 0$  and $\text{tr}\left(\lambda_i \lambda_j\right) = 2 \delta_{ij}$. Therefore, we can express any density matrix $\rho$ as
\begin{align}
\rho=\frac{1}{n}\left(\mathbb{I}+\sqrt{\frac{n(n-1)}{2}}\mathbf{x}\cdot\vec{\lambda}\right),
\end{align}
where \(\mathbf{x}=(x_1,x_2,\dots,x_{n^2-1})\in \mathbb{R}^{n^2-1}\)
satisfies
\begin{align}
x_i = \sqrt{\frac{n}{2(n-1)}}\text{tr}\left(\rho \lambda_i\right).
\end{align}

Thus, each density matrix can be represented as eight real numbers. The input data for the CNN are density matrices (as real numbers) on a $10 \times 10 \times 10$ momentum grid. For each $h$, we numerically generate the input sample of density matrices. The entire data set consists of 5000 samples, where $h$ is varied uniformly from $h = -5$ to $h =5$. Removing the interval $[1.8,2.2]$ and $[3.8,4.2]$, where the experimental data lie, we are left with 4601 samples. This is further divided into the training set and validation set randomly with a ratio of $3:1$, producing 3680 training samples and 921 validation samples.

The CNN architecture consists of three layers: a 3D convolution layer, a 3D maxpooling layer, a dropout \cite{srivastava2014dropout} layer with rate $0.4$, and a fully-connected flattening layer with $0.55$ dropout regularization. The last layer is then connected to the final softmax classifier, which outputs the probability for each possible topological phase. In our case, we have three categories, which are labelled as $\theta^{(m)}/\pi = 0,1,-2$. The dropout layer avoids overfitting and is crucial for maintaining good performance when parts of the data are missing. We choose the RMSprop optimizer \citep{Tieleman2012} and a batch size of 128. The learning rate is set to be $10^{-3}$ during training.

\subsection{Learning phase transitions}

\begin{figure}[b]
\begin{center}
\includegraphics[trim=0cm 0cm 0cm 0cm, clip,width=\linewidth]{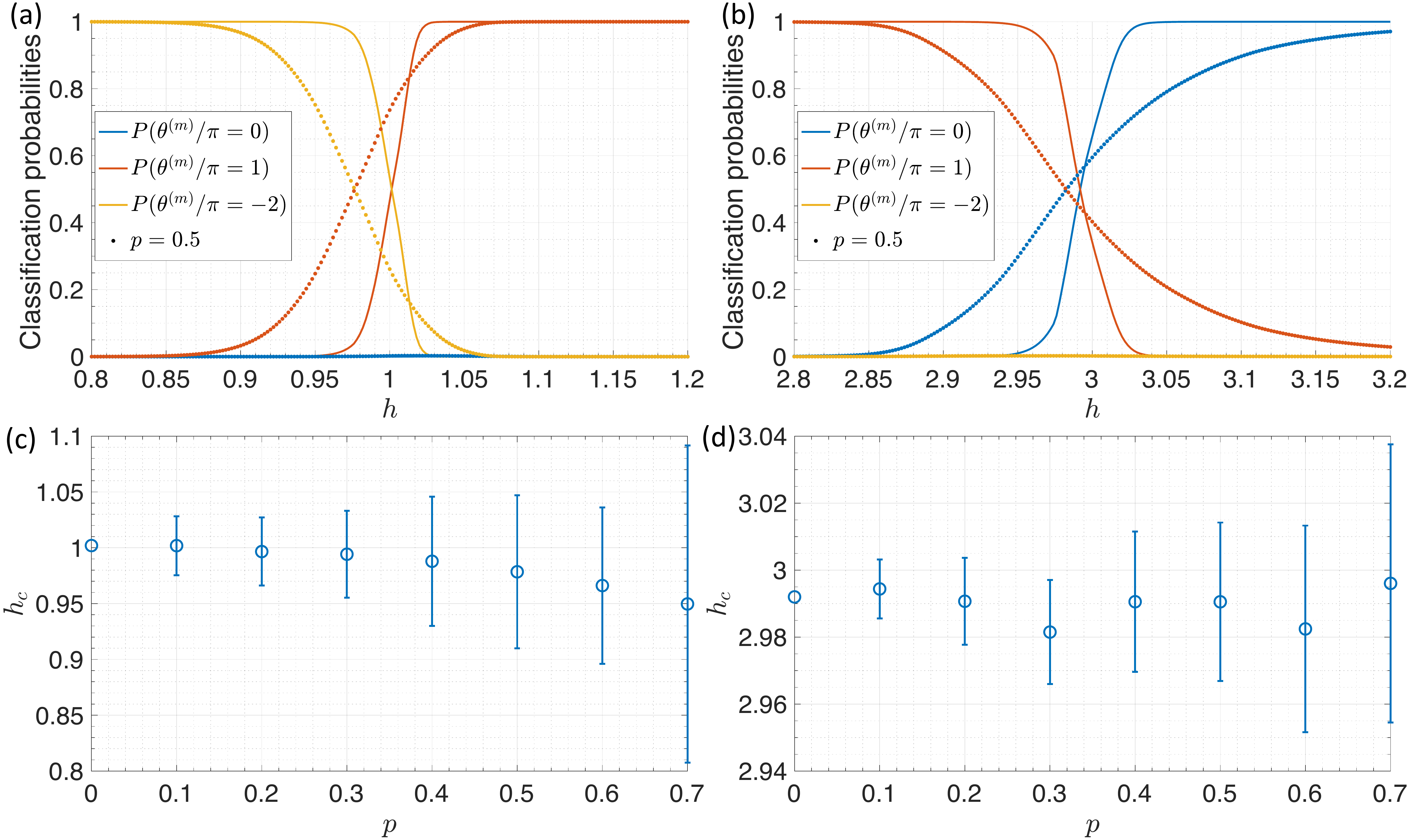}
\caption{Learning topological phase transitions. The CNN is trained with correct labels of topological phases from $h = -5$ to $h =5$, except for the intervals $[-3.5,-2.5]$, $[-1.5,-0.5]$, $[0.5,1.5]$, and $[2.5,3.5]$. The trained CNN is then applied to learn the topological phase transition points. (a) Topological phase transition near $h = 1$. The solid curves denote the prediction probabilities for respective phases with the input as the full density matrix data on a $10 \times 10 \times 10$ grid. The dots are corresponding prediction probabilities with partial data input, where a proportion $p=0.5$ is randomly discarded and each dot is averaged over 100 random realizations. (b) Similar to (a) but for phase transitions near $h = 3$. (c) Learning the topological phase transition point near $h = 1$. For each $p$ (except $p=0$), the phase transition point $h_{c}$ is averaged over 100 random realizations, with standard deviations giving rise to the error bar. (d) Learning the topological phase transition points near $h = 3$.}
\label{Fig:SM_transition}
\end{center}
\end{figure}

In this subsection, we demonstrate that CNN can be used to learn topological phase transitions. The use of machine learning tools to study phase transitions is pioneered by Ref.~\cite{Carrasquilla2017Machine, Nieuwenburg2017Learning, Chng2017Machine}. Here, we adapt the idea to topological phase transitions. In principle, the entire training and learning process can be performed with purely experimental data. However, due to the extremely long data taking time for various $h$ values, we use numerical simulation here to demonstrate the power of the machine learning approach. 

The basic idea is the following: suppose the distinct phases of matter can be characterized (labelled) far away from the phase transition boundaries; the correct labels of the phases can be used to train a CNN, which can subsequently be applied to study phase transition behavior and identify phase transition points. This is the common scenario when one studies phase transitions. For example, we can easily identify and distinguish paramagnetic phases and ferromagnetic phases far away from the phase transition boundaries. The CNN can then be trained and applied to explore the phase transition behavior. 

Here, we demonstrate the above capabilities of the machine learning approach. We first train a CNN with correct labels of topological phases from $h = -5$ to $h = 5$, except for the intervals $[-3.5,-2.5]$, $[-1.5,-0.5]$, $[0.5,1.5]$, and $[2.5,3.5]$. We know topological phase transitions occur in those parameter ranges and would like to use a CNN to identify phase transition points. Fig.~\ref{Fig:SM_transition}(a) shows the prediction of phases near $h = 1$. With the full set of data on a $10 \times 10 \times 10$ grid, we can identify a sharp phase transition point $h_{c}$, where the probabilities cross. If we remove a $p = 0.5$ proportion of the data, the phase transition becomes less sharp, and the transition point also shifts further away from the theoretical value ($h = 1$). Fig.~\ref{Fig:SM_transition}(b) shows a similar plot for phase transition near $h = 3$. 

We locate the phase transition point from the crossing point of prediction probabilities. Fig.~\ref{Fig:SM_transition}(c) and Fig.~\ref{Fig:SM_transition}(d) plots the phase transition point $h_{c}$ versus the data dropping proportion $p$. For each $p$ (except $p=0$), the phase transition point $h_{c}$ is averaged over 100 random data dropping realizations. We can clearly see that as we use less data in the input ($p$ increases), accuracies decrease and uncertainties grow. We note that quantum fluctuations increase near phase transitions, so more data are needed for the correct recognization of the phases. This is in contrast with Fig.~2(c) of the main text, where $h=2$ and $h=4$ are far from the phase transition, so those phases can be learnt correctly with less data.

\section{Conventional method to measure topological invariants}

\begin{figure}[b]
\begin{center}
\includegraphics[trim=0cm 0cm 0cm 0cm, clip,width=0.7\linewidth]{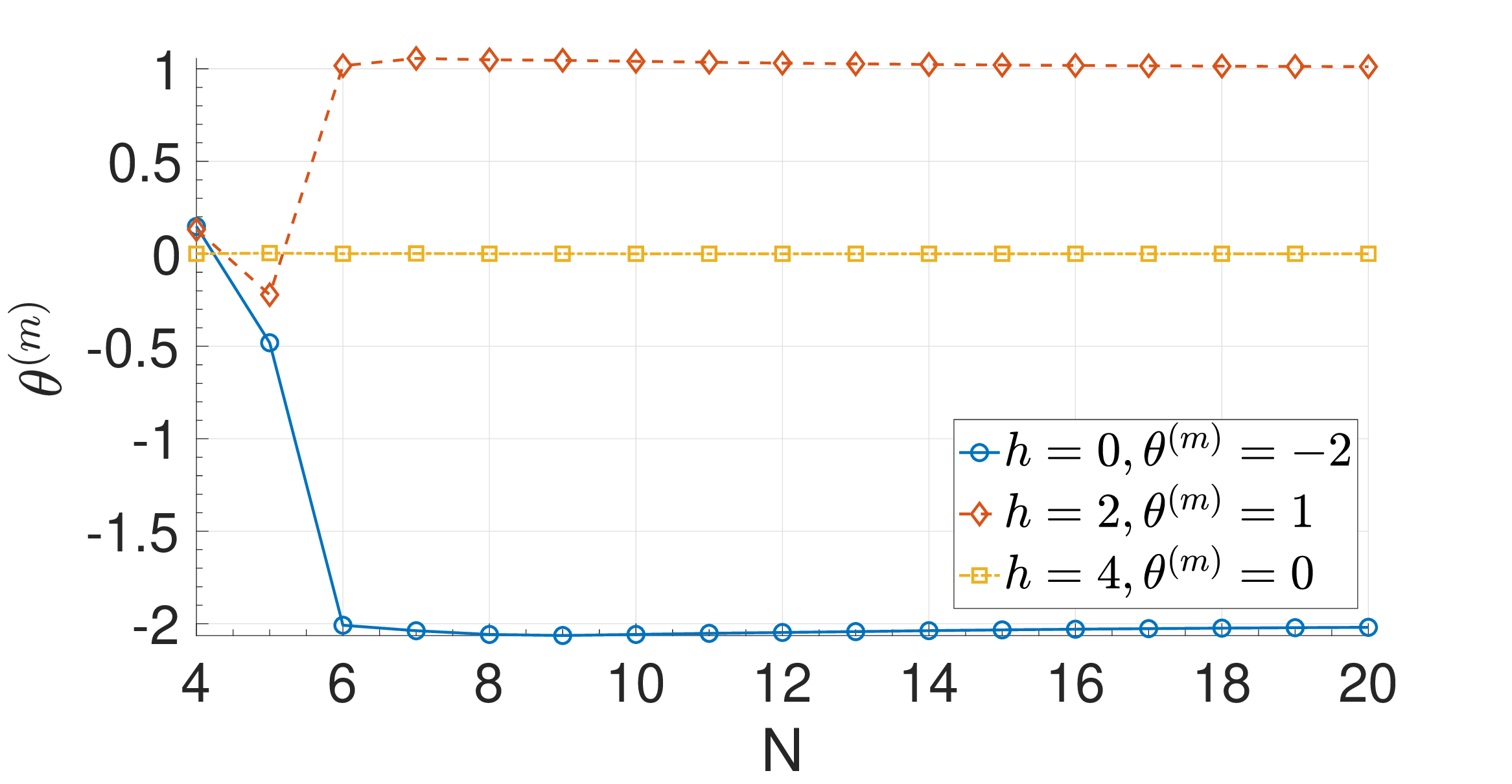}
\caption{Theoretically extracted values of the topological invariant $\theta^{(m)}$ based on the discretization method from a $N \times N \times N$ momentum grid.}
\label{Fig:TopInvariant_sm}
\end{center}
\end{figure}

\begin{table}[t]
\caption{Measured topological invariant $\theta$ for the middle band based on the discretization method. The topological invariant is extracted from experiment by full quantum state tomography on an equally-spaced $N \times N \times N$ momentum grid. Uncertainties in the experimental values are obtained from Monte Carlo simulations by assuming a Poissonian distribution for the measured fluorescence counts. Corresponding theoretical values on a finite and an infinite grid are shown in comparison.}
\label{Table:TopInv}

\begin{ruledtabular}
\begin{tabular}{cccc}
$h$ & \multicolumn{3}{c}{$\theta^{(m)}/\pi$}
\\
\cline{1-1} \cline{2-4}
 & Experiment ($N=10$) & Theory ($N=10$)  & Theory ($N=\infty$)
 \\
2 & 1.032(4) & 1.041 & 1
\\
4 & $-0.0015(5)$ & $4\times 10^{-6}$ & 0
 \\
\end{tabular}
\end{ruledtabular}
\end{table}

In this work, we mainly used the convolutional neural network to learn the topological invariants. Alternatively, one can also make use of the state-of-the-art discretization method based on quantum state tomography proposed in Ref.\ \cite{Fukui2005Chern, Deng2014Direct} and implemented recently in Ref.\ \cite{Yuan2017Observation} for measuring the Hopf invariant. The scheme is robust to finite discretization error and is able to overcome the problem of local gauge ambiguity. Table \ref{Table:TopInv} shows the measured topological invariant compared to the theoretical value from state tomography on a $10 \times 10 \times 10$ momentum mesh grid. Theoretically, we can also simulate the measurement of topological invariants on a smaller mesh grid, as shown in Fig.~\ref{Fig:TopInvariant_sm}. This scheme is quite robust and correct values can be extracted for $N \geq 6$. 

Here, we make a few comments on comparing the machine learning approach and the conventional method. First, the CNN does not need to know the formula for topological invariants, so it has the power to be generalized to less well-understood phases of matter, where the order parameters or topological invariants are not a priori known. In addition, the machine learning tools have the advantage of directly using measurement results in the form of mixed states. In terms of data efficiency, CNN is slightly more efficient and robust, though a direct comparison is not possible. The discretization method used here does not find the correct topological invariants on a $5\times 5 \times 5$ grid, but the CNN can still identify the correct phases with 5\% data on a $10 \times 10 \times 10$ grid with $\sim 80\%$ probability. In addition, the neural network still works surprisingly well even for a small number of randomly distributed data samples, whereas the extraction of topological invariants based on the method here may not be straightforward when measurement data are spread over random momentum points.

\section{Energy spectrum}

\begin{figure}[b]
\begin{center}
\includegraphics[trim=0cm 0cm 0cm 0cm, clip,width=\linewidth]{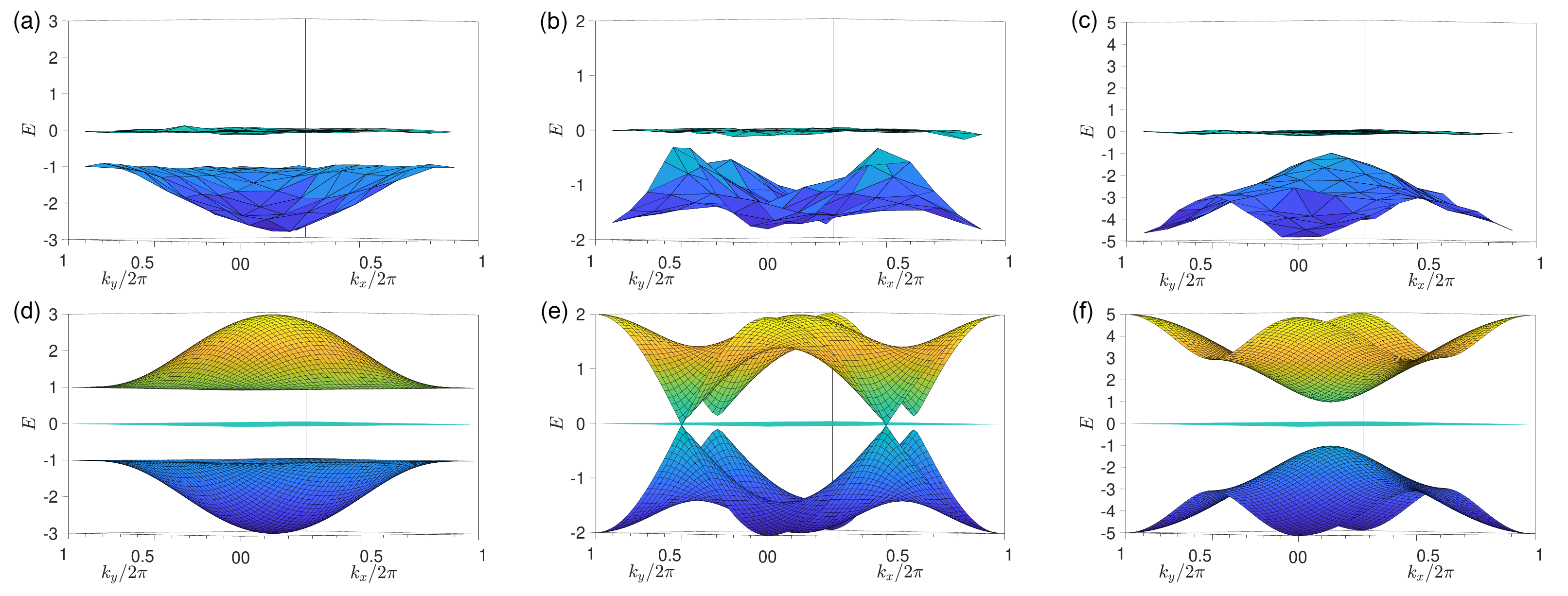}
\caption{Experimentally measured energy dispersion of each band for the momentum layer $k_{z}=\pi$ with (a) $h = 0$, (b) $h = 1$, and (c) $h=4$, together with the corresponding theoretical plots in (d),(e), and (f). Energies are measured via quantum state tomography on a $10 \times 10$ grid in the $(k_{x},k_{y})$ plane, and the surface plots are interpolated from the grid.}
\label{Fig:ES}
\end{center}
\end{figure}

In Fig.~2 of the main text, we plot the energy spectrum for $h = 2$ and $ h = 3$. Here, we include, in Fig.~\ref{Fig:ES}, additional plots for $ h = 0,1, 4$ to further probe the dispersionless middle band as well as the change in lower band. Again, we measure energy spectrum on a $10 \times 10$ grid in the $(k_{x},k_{y})$ plane for the momentum layer $k_{z} = \pi$. For $h = 0$ and $h = 4$, we can clearly see a band gap, while for $h = 1$, the gap closes at $(k_{x},k_{y},k_{z}) = (0, \pi, \pi)$ and $(\pi, 0, \pi)$, which is less obvious due to finite resolution.